\documentclass[journal,10pt]{IEEEtran}

\usepackage{amssymb}
\usepackage{amsmath}
\usepackage{cite}
\usepackage{url}
\usepackage{xcolor}
\usepackage{graphicx,amsmath,amssymb}
\usepackage{subfigure}
\usepackage{citesort}
\usepackage{fancyhdr}
\usepackage{mdwmath}
\usepackage{mdwtab}
\usepackage{amsthm}
\usepackage{lineno}
\usepackage{threeparttable}
\usepackage{algorithm}
\usepackage{algorithmic}
\usepackage{multirow}
\usepackage{flafter}

\newcommand{\tabincell}[2]{\begin{tabular}{@{}#1@{}}#2\end{tabular}}

\makeatletter
\def\ScaleIfNeeded{%
\ifdim\Gin@nat@width>\linewidth \linewidth \else \Gin@nat@width
\fi } \makeatother

\begin{document}
\title{Developing NOMA to Next Generation \\Multiple Access (NGMA): \\Future Vision and Research Opportunities}

\author{
 Yuanwei~Liu, Wenqiang~Yi, Zhiguo Ding, Xiao Liu, Octavia Dobre, and Naofal Al-Dhahir
}

\maketitle

\begin{abstract}
  As a prominent member of the next generation multiple access (NGMA) family, non-orthogonal multiple access (NOMA) has been recognized as a promising multiple access candidate for the sixth-generation (6G) networks. This article focuses on applying NOMA in 6G networks, with an emphasis on proposing the so-called ``One Basic Principle plus Four New'' concept. Starting with the basic NOMA principle, the importance of successive interference cancellation (SIC) becomes evident. In particular, the advantages and drawbacks of both the channel state information based SIC and quality-of-service based SIC are discussed. Then, the application of NOMA to meet the new 6G performance requirements, especially for massive connectivity, is explored. Furthermore, the integration of NOMA with new physical layer techniques is considered, followed by introducing new application scenarios for NOMA towards 6G. Finally, the application of machine learning in NOMA networks is investigated, ushering in the machine learning empowered NGMA era.
\end{abstract}

\section{Introduction}
\begin{figure*}[t]
  \centering
    \includegraphics[width= 0.7 \linewidth]{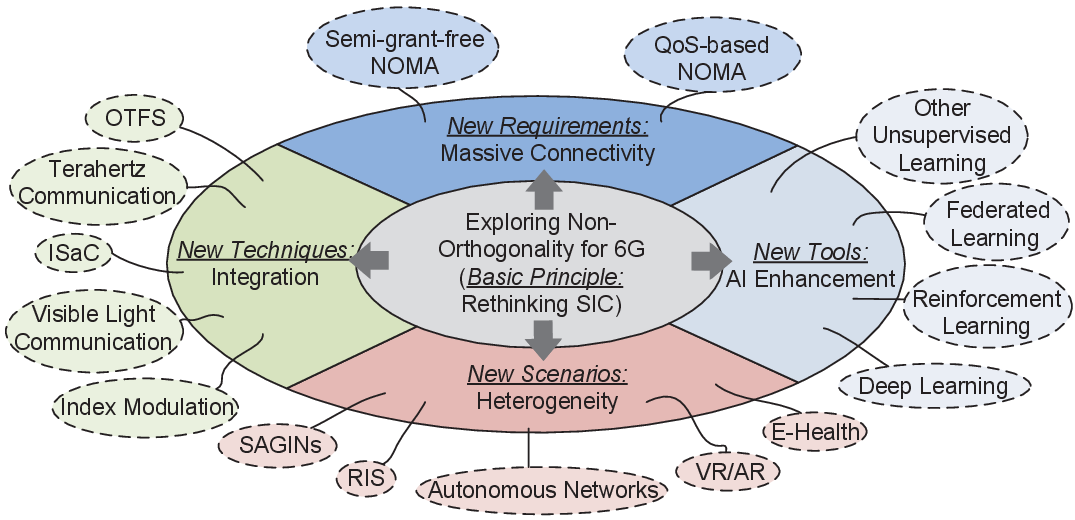}
  \caption{Multiple access techniques for 6G by exploring non-orthogonality with the ``One Basic Principle plus Four New'' concept, where OTFS, ISaC, and RIS represent orthogonal-time-frequency-space, integrated-sensing-and-communication, and reconfigurable-intelligent-surface, respectively. }\label{sys}
\end{figure*}
According to the latest Visual Network Index report~\cite{report}, the global mobile data traffic in 2022 will be seven-fold larger compared to 2017. In addition to this explosive data demand, emerging bandwidth-thirsty applications, e.g., space-air-ground-integrated-networks (SAGINs), virtual reality (VR), augmented reality (AR), etc., requiring strict quality-of-service (QoS) create further challenges for next-generation wireless networks.

To solve these challenges, the sixth-generation (6G) networks need breakthroughs beyond the current fifth-generation (5G) networks~\cite{A2}. The expected performance targets of 6G are: 1) The connectivity density is ten-fold larger compared to 5G; 2) The peak data rate reaches 1 terabit per second; 3) The energy efficiency is a hundred times higher than that of 5G; 4) The air interface latency decreases to 0.1 millisecond; and 5) The reliability increases to 99.99999\%. In addition, unlike 5G focusing on a single objective, most 6G scenarios need to optimise multiple objectives simultaneously~\cite{A1}. To this end, highly efficient next-generation multiple access (NGMA) techniques are vital for 6G.

Due to introducing an additional degree of freedom in the power domain, non-orthogonal multiple access (NOMA) has attracted significant attention from both academia and industry. For decoding different signals in the same orthogonal (time/frequency) resource block (ORB), NOMA employs superposition coding at the transmitters and successive interference cancellation (SIC) at the receivers. Therefore, NOMA is able to provide additional access for overloaded traffic scenarios, which is a common scenario in 6G with heterogeneous ultra dense networks~\cite{T_NOMA}. Although NOMA has already been thoroughly investigated in the 5G and beyond networks, previous research focused on static devices and the data rate of broadband users. This ignores several fundamental problems for NGMA, e.g., the effect of mobility, the design tradeoffs in terms of connectivity, reliability, and latency.

Our goal in this article is to fill this gap by investigating NOMA via the ``One Basic Principle plus Four New'' concept as illustrated in Fig.~\ref{sys}. More specifically, the basic principle is to deeply investigate the non-orthogonality from the SIC perspective, which is rarely discussed in existing NOMA-related magazines. Building on this basic principle, we explore the following four new directions: 1) New requirements: Supporting massive connectivity by considering various QoS requirements including latency and reliability; 2) New techniques: Integrating NOMA with other 6G physical layer techniques; 3) New applications: Application in heterogeneous scenarios to support emerging 6G applications; and 4) New tools: Integration with artificial intelligence (AI) to design an adaptive resource allocation.

\section{Basic Principle: Rethinking SIC in NOMA}
\begin{figure}[t]
  \centering
  \includegraphics[width= 1 \linewidth]{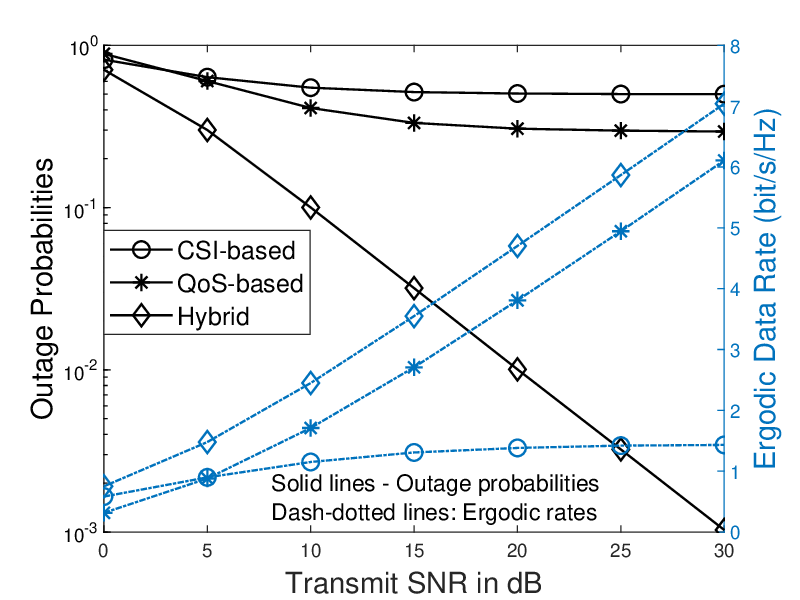}\\
  \caption{Illustration of the performance gain of hybrid SIC decoding order over conventional schemes~\cite{hybrid_SIC}.}\label{fig1}
\end{figure}
The SIC process is the key component used by various forms of NOMA, and for these conventional NOMA, SIC decoding has been designed by using a single metric, which is to ensure low implementation complexity. Take power-domain uplink NOMA as an example, where the channel state information (CSI) is used as a metric for determining SIC decoding order, i.e., users are distinguished by their channel conditions. Without loss of generality, consider a special case with a strong user and a weak user. The CSI-based SIC decodes the strong user's signal first. If the first stage of SIC is successful, the base station (BS) then decodes the weak user's signal. The CSI-based SIC is an intuitive decoding strategy and it is able to effectively explore the users' dynamic channel conditions. Another implementation of SIC is based on the users' QoS requirements, as in cognitive-radio inspired NOMA. Consider a two-user case as an example. Assume that the primary user is an Internet-of-Things (IoT) sensor and the secondary user is a broadband user. The QoS-based SIC detects the primary user first, because the data rate achievable during the first SIC stage can be extremely low, due to strong co-channel interference. Therefore, it is reasonable to first decode the signal of the sensor which is to be served with a low data rate.

While these conventional SIC decoding schemes    can be implemented with low complexity and have their own advantages in different application scenarios, it is important to point out that they are not optimal. For example, both the CSI- and QoS-based SIC decoding orders   suffer from the same drawback, i.e., an outage probability error floor is inevitable, as explained in the following. For CSI-based SIC, at high signal-to-noise ratio (SNR), the strong user's signal-to-interference-plus-noise ratio (SINR) becomes a constant. This means that the outage probability for the first SIC stage will never go to zero, regardless of how large the user's transmit power is. Following a similar reasoning, one can also conclude that QoS-based SIC also suffers from this outage probability error floor which cannot be avoided by simply increasing the users' transmit power levels. Surprisingly, this `inevitable' floor can be avoided by using a simple SIC decoding strategy, termed hybrid SIC, as discussed in~\cite{hybrid_SIC}. In particular, the hybrid SIC takes both the users' QoS requirements and their channel conditions into consideration. Again, consider the simple case of one primary user and one secondary user as an example. Unlike CSI-based NOMA which compares the two users' channel conditions, the hybrid SIC evaluates the quality of a user's channel condition by taking the users' QoS requirements into consideration. For example, if the secondary user's channel is strong enough to guarantee the primary user's QoS requirement, the secondary user's signal is decoded during the first SIC stage. Otherwise, the primary user's signal is decoded first. The performance gain of the hybrid SIC over conventional SIC schemes is shown in Fig.~\ref{fig1}, which demonstrates that  the outage probability error floor suffered by the two conventional schemes can be avoided. It is important to point out hybrid NOMA is still not optimal, and an important  direction for future research is to design more sophisticated  SIC for optimizing  the reception reliability and data rates of NOMA transmission.

\section{New Requirements: Massive Connectivity for NOMA}
As a multiple access technique, the most important task for NGMA in 6G is to provide massive connectivity. Due to obtaining diverse QoS requirements in most 6G applications, the massive connectivity provided by NGMA needs to be constrained by different QoS requirements in terms of user experience, which include not only data rates but also latency and reliability. After that, NGMA is capable of maximising the spectral efficiency of 6G networks. It worth mentioning that a high spectral efficiency also contributes to energy conservation for 6G. To satisfy these performance metrics and hence guarantee massive connectivity, the design of downlink NOMA becomes QoS-oriented, and semi-grant-free (semi-GF) transmissions are preferable for uplink NOMA.

\subsection{QoS-based NOMA for Downlink Transmission}
Note that multiple SIC iterations introduce long waiting times and high outage probabilities. For downlink NOMA in 6G, the main challenge is to overcome high latency and low reliability due to SIC, especially for the massive users scenario. In addition to applying the hybrid SIC for enhancing the spectral efficiency, QoS-based NOMA needs to design principles of QoS-based user clustering (Q-UC) and QoS-based power allocation (Q-PA) in terms of connectivity.

For Q-UC, two fundamental questions should be answered: 1) Which is better, grouping the same or different QoS users in a NOMA cluster? 2) How to avoid SIC for latency-reliability-sensitive users? For better illustration, we use broadband users (aiming to maximize data rates) and IoT sensors (requiring low latency and high reliability) to depict two typical QoS types of IoT users in 6G. Regarding the first question, there is no single correct answer, but in general, each NOMA cluster favors including different QoS users since users with the same QoS compete in the same domain. Regarding the second question, it is obvious that NOMA favors decoding the IoT sensor first to reduce the SIC-caused latency. Therefore, each IoT sensor prefers broadband users with stronger channel gains than itself.

For Q-PA, the desired power for IoT sensors needs to consider the max-rate between the successful decoding rate (SDR) and the latency-reliability-based data rate, while that for the SIC process of broadband users just needs the SDR. Note that the conventional NOMA scheme only needs SDRs for power allocation. Moreover, to maximize the connectivity, the allocated power only needs to satisfy the QoS requirement of each user.

\subsection{Semi-GF NOMA for Uplink Transmission}
Uplink NOMA in 6G needs to serve massive intermittently active users whose number is much larger than the available ORBs. To solve this problem, a promising approach is to group grant-based (GB) users with GF users in one NOMA cluster, which is defined as Semi-GF NOMA~\cite{SGF}. The general process of Semi-GF NOMA is as follows: 1) The BS grants one ORB to each GB user; 2) The BS evaluates the power resource left, i.e., the acceptable received power from GF users, for this ORB and then calculates a threshold for broadcasting; and 3) After receiving the broadcast threshold, the satisfied GF users transmit their messages to the BS via GF transmissions. Based on whether the maximum load in each ORB is available or not, the threshold design can be divided into single-ORB and multi-ORB cases.

For the single-ORB case, it can be assumed that the ORB for each GF user is pre-allocated and, hence, the number of potential GF users (PGFU) in each ORB is known in advance. If the GB user is delay-tolerant, several SIC iterations are acceptable. Therefore, the broadcast threshold contains two kinds of information: 1) Minimal transmit power (MTP); and 2) Maximal tolerance to interference (MTI). If GF users have transmit power higher than MTP, they upload messages and the successful GF users are decoded before the GB user. To avoid frequent collisions, the MTP-threshold is suitable for GF users with a relatively low activation rate. If GF users have transmit power lower than MTI, they are also able to upload messages that are decoded after the GB user. Note that the MTI needs to take all PGFUs into account to avoid the transmission outage of the GB user. This restriction can be relaxed when the GB user is based on long-term communications. In this case, the MTI only needs to consider the average number of active GF users, which is significantly smaller than the total number of PGFUs. If the GB user is delay-sensitive, it needs to be decoded firstly, so the threshold only contains the MTI.

\begin{figure}[t!]
\centering
\includegraphics[width=1 \linewidth]{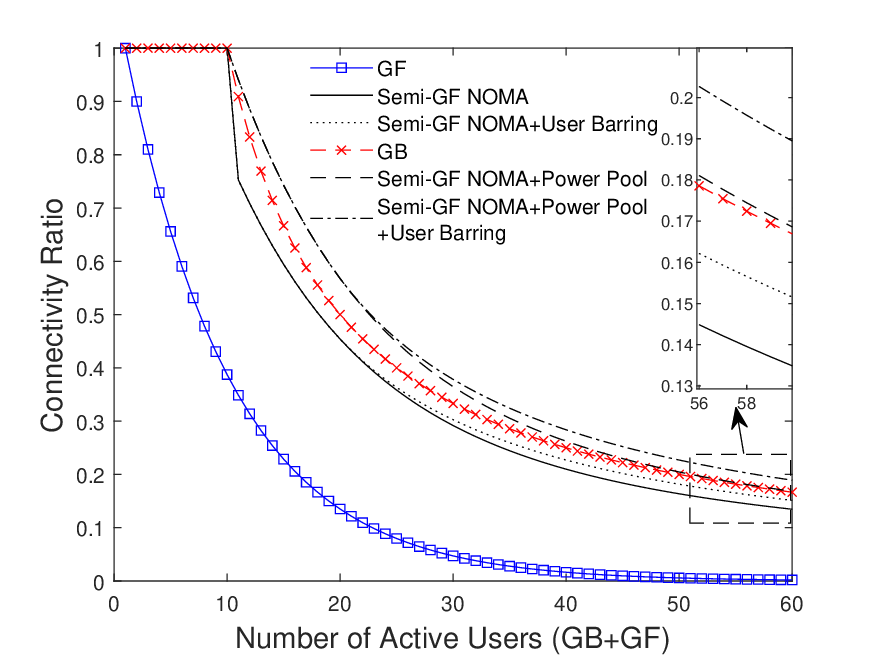}
\caption{Connectivity performance of semi-GF NOMA with $10$ ORBs, where each NOMA cluster supports maximum two users\cite{fayaz2020transmit,yu2020}.}\label{Fig1b}
\end{figure}

For the multi-ORB case, the number of PGFUs for each ORB is unknown since GF users are capable of choosing ORBs at random, which results in a complicated threshold design. To reduce collisions, it is better to control the transmit power of Semi-GF NOMA. Based on channel gains, a power pool design with layered transmit power levels~\cite{fayaz2020transmit} can be applied to decrease outage probabilities. By restricting the number of GF users in each ORB, an access class barring (ACB)~\cite{yu2020} aided user barring approach can be utilized to further enhance the connectivity. However, the protocol design of semi-GF NOMA under multi-ORB cases still faces numerous challenges. As illustrated in Fig.~\ref{Fig1b}, the connectivity of semi-GF NOMA with ideal power pool and user barring is the highest. Without controlling the transmit power, semi-GF NOMA experiences more collisions than traditional GB transmission.

\section{New Techniques: Integration of NOMA with Emerging Physical Layer Techniques}
In 6G, numerous emerging techniques for the physical layer will be gradually developed to address fundamental problems in future networks. For example, the orthogonal-time-frequency-space (OTFS) modulation aims to enhance the performance in a challenging communication scenario with high mobility. Terahertz (THz) communications are able to provide multi-GHz bandwidth to address the lack of spectrum resources. The integration of NOMA with OTFS, THz, and other promising techniques is an exciting research challenge for 6G.

\subsection{OTFS-NOMA}
For traditional slow-mobile (or static) users, their channels in a coherence time can be regarded as invariable. With the aid of orthogonal frequency-division multiplexing (OFDM), the communication performance can be evaluated in the time-frequency plane with ORBs. However, the communication quality of fast-mobile users is mainly decided by the delay and Doppler effects. It is worth pointing out that the traditional OFDM subcarriers in the delay-Doppler plane are not orthogonal, which introduces strong inter-symbol interference and inter-carrier interference for fast-mobile users. To solve this problem, a new modulation scheme, referred to as OTFS~\cite{OTFS-NOMA}, was proposed to provide nearly non-fading channels for these doubly dispersive channels.

Since the channel state for fast-mobile users is commonly worse than that of slow-mobile users, NOMA can be an efficient technique by exploiting this channel disparity. In this part, we discuss a general scenario that the slow-mobile user has a stronger channel than the fast-mobile user. In OTFS-NOMA systems, one slow-mobile user and one fast-mobile user can be grouped into a NOMA cluster. As the slow-mobile user has a strong channel, the SIC process is carried out by this user. Unlike conventional NOMA systems, the SIC at the slow-mobile user has two unique stages: 1) The slow-mobile user first decodes the fast-mobile user's signal in the delay-Doppler plane; and 2) After cancelling the decoded signal in the first stage from the received signal, the slow-mobile user decodes its signal in the time-frequency plane. For the fast-mobile user, it decodes its own signal directly in the delay-Doppler plane. The proposed OTFS-NOMA scheme benefits both the fast- and slow-mobile user in terms of latency and
spectral efficiency. More specifically, fast-mobile users obtain additional ORBs to enhance the performance and slow-mobile users have the opportunity to access spectrum resources that are solely allocated to fast-mobile users under the conventional OTFS modulation with orthogonal multiple access (OMA).
\begin{figure}[t!]
  \centering
    \includegraphics[width= 1 \linewidth]{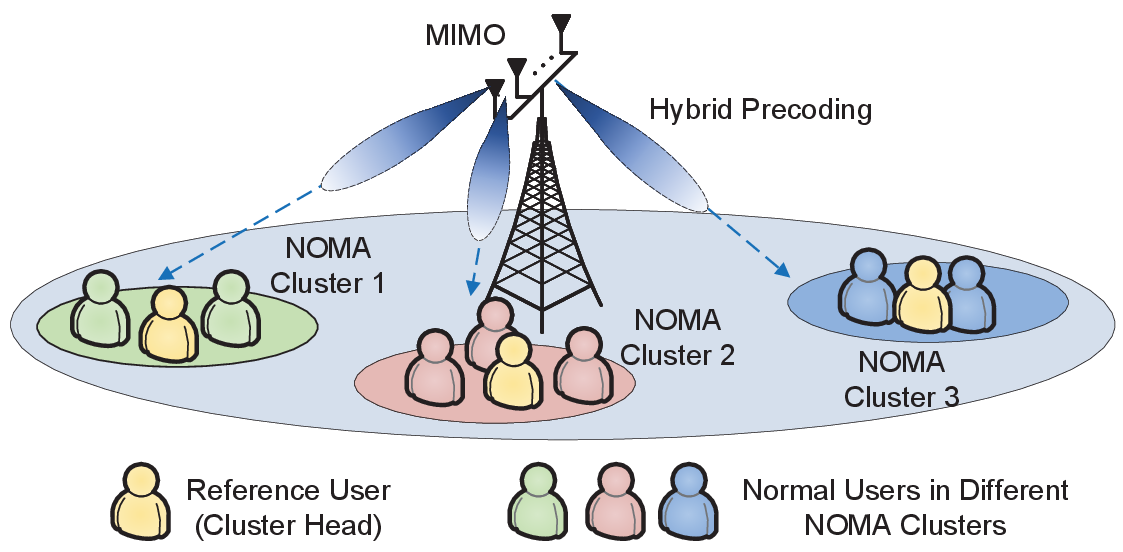}
  \caption{Illustration of location-based user clustering in THz-NOMA systems.}\label{Fig2}
\end{figure}
\subsection{THz-NOMA}
THz communication is a highly-promising technology for 6G, whose bandwidth is 3-4 orders of magnitude larger than current wireless systems. Due to its short wavelength, super massive multiple-input multiple-output (SM-MIMO) becomes an indispensable component for THz communications to exploit the spatial diversity. In general, if the number of users with independent channels is much smaller than the number of antennas, the channels between different users are nearly orthogonal. In this case, NOMA is not a more efficient multiple access technique than traditional solutions for multi-user MIMO (MU-MIMO) systems. However, the low-rank channels of THz users are highly correlated because of the limited-scattering transmission. Therefore, NOMA becomes a promising technique to improve the spectral efficiency in THz aided networks.

For THz communications, NOMA aims to group all highly correlated users into one NOMA cluster and allocate to them different transmit power levels for decoding according to the hybrid SIC as mentioned in Section II. After that, each NOMA cluster has one beam and different NOMA clusters are served by MU-MIMO techniques. As shown in Fig.~\ref{Fig2}, since the line-of-sight link is the most important transmission path in THz communications, grouping users located in the same geographic area into one cluster becomes possible. This user clustering design has low complexity since it only needs the full CSI of one reference user, i.e., cluster head, in each NOMA cluster. A recent work~\cite{THz-NOMA} proposed a machine learning aided solution for THz-NOMA based on a hybrid precoding scheme with a sub-connection structure, where the K-means machine learning algorithm is used for user clustering and a distributed alternating direction method of multipliers algorithm is applied for power allocation. Based on this solution, the energy efficiency of THz-NOMA systems can be maximized. This interesting work can be further extended by considering location-based user clustering to reduce the channel estimation complexity.

\subsection{Integration of NOMA with Other Techniques}
In addition to OTFS-NOMA and THz-NOMA, other promising physical layer techniques will also benefit from NOMA. Some typical use cases for NGMA systems in 6G are listed as follows:
\begin{itemize}
     \item \textbf{Integrated Sensing and Communication (ISaC):} NOMA is capable of helping receivers to distinguish communication and radar-return signals. The communication signal can be decoded first and cancelled from the received signal via SIC. Then, the radar signal can be analyzed via conventional sensing algorithms.
     \item \textbf{Visible Light Communication (VLC):} NOMA is able to provide high spectral efficiency for VLC to solve the small modulation bandwidth problem of light-emitting diodes, especially for multi-user cases. NOMA also benefits from the high SNR offered by VLC.
     \item \textbf{Index Modulation (IM):} NOMA is capable of handling the strong inter-user interference issue in IM. The users with similar spatial features can be grouped into one NOMA cluster, hence increasing both spectral efficiency and energy efficiency.
\end{itemize}

\section{New Scenarios: Application of NOMA to Heterogeneous Scenarios}

6G networks are expected to be transformative and revolutionary encompassing applications like data-driven, ubiquitous wireless connectivity, and connected intelligence. Although we are still in the early stage to define 6G networks, some scenarios with heterogeneous traffics are recognized as potential candidates towards 6G. This section discusses the application of NOMA by considering the basic principle to new scenarios towards 6G~\cite{Zhang2018Energy}.

\begin{figure*}[t!]
  \centering
    \includegraphics[width= 0.8 \linewidth]{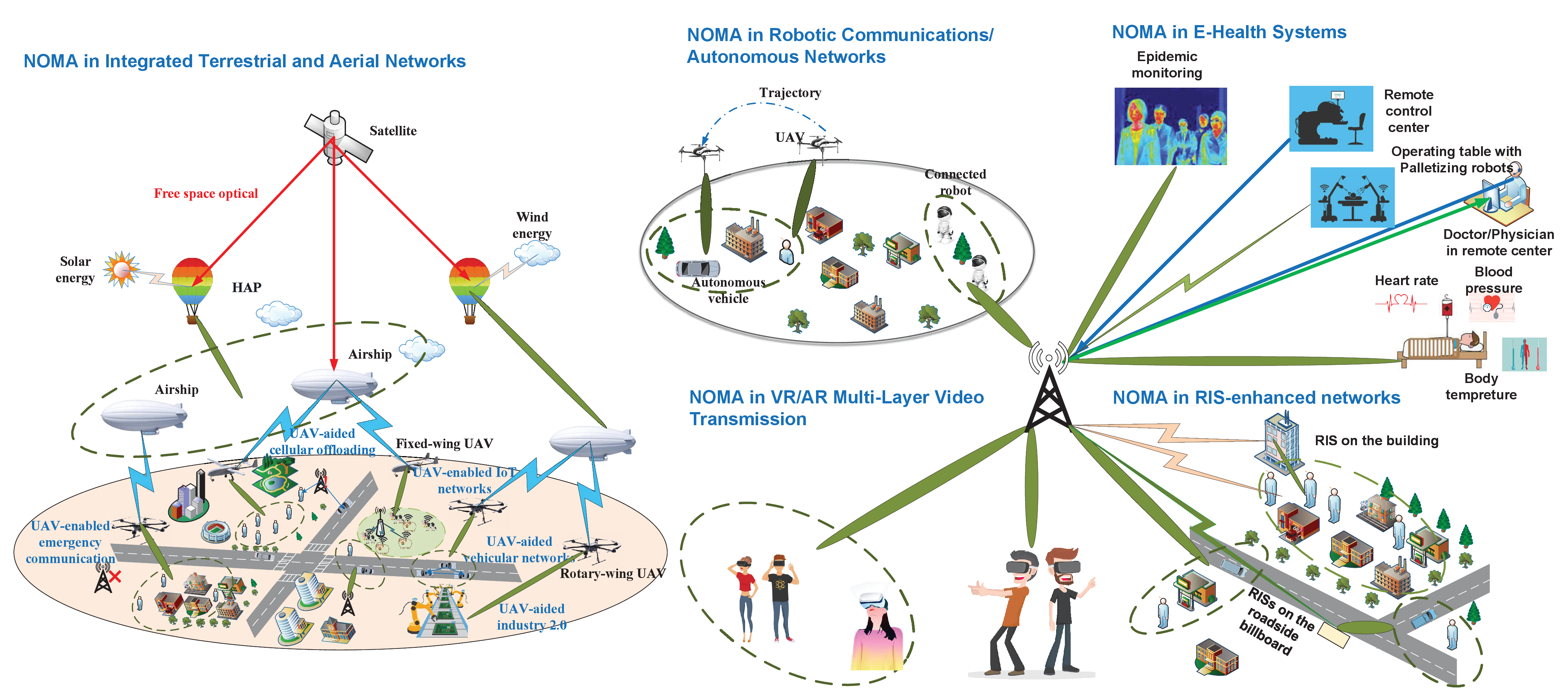}
  \caption{Application of NOMA in scenarios towards 6G.}\label{application}
\end{figure*}
\subsection{NOMA in Integrated Terrestrial and Aerial Networks}
SAGINs, which intrinsically integrate unmanned aerial vehicle (UAV)-aided terrestrial networks with sky-platforms and satellites, have become a focal point in wireless communications research as they address a wide range of challenges encountered in single-tier networks. As shown in Fig.~\ref{application}, based on the cooperation of high altitude platforms (HAPs) and low altitude platforms (LAPs), drones, aircraft, and satellite, the conceived multi-tier SAGINs are capable of filling the coverage-holes of single-tier networks by integrating the advantages of each network. In contrast to terrestrial wireless networks, the challenging optimization problems encountered in SAGINs have to provide massive connectivity under multiple objectives (delay, throughput, bit error rate (BER), power) to arrive at an attractive solution. For example, in the case of low earth orbit systems, a large number of satellites are needed to cover the globe. SAGINs rely on the seamless integration of heterogeneous network segments with the goal of providing uninterrupted and ubiquitous connectivity to everyone, everything, and everywhere. As mentioned in Section II, the hybrid SIC enabled NOMA can be applied in SAGINs to enhance not only data rates but also delay and BER, since hybrid SIC is promising to meet heterogeneous QoS requirements in multi-tier SAGINs while guaranteeing ubiquitous and massive connectivity.
\subsection{NOMA in Reconfigurable-Intelligent-Surface-Enhanced Networks}
Owing to their ability of smartly reconfiguring the wireless propagation environment via passive reflecting elements, reconfigurable intelligent surfaces (RISs) are capable of providing potential received SINR enhancements by constructing a software-defined wireless environment. Compared to orthogonal multiple access (OMA) RIS networks, NOMA-RIS networks can achieve higher spectral efficiency and enhanced massive connectivity. More importantly, NOMA-RIS networks are more flexible than the conventional power-domain NOMA-based networks because RISs can improve or degrade the channel quality of individual users by adjusting the phase shifts of reflecting elements and their positions which, in turn, smartly controls the channel conditions of users. Additionally, by integrating RISs in MIMO-RIS networks, the strict constraint on the number of antennas at the transmitters and receivers can be relaxed due to the multiple reflecting elements on RISs. For RIS-enabled networks, traditional CSI-based SIC brings high implementation complexity since perfect CSI for cascaded channels at the transmitter is desired. Fortunately, QoS-based SIC can be used to relax this limit. By integrating hybrid-SIC into NOMA-RIS networks is able to handle the tradeoff between channel estimation complexity and high spectral efficiency.
\subsection{NOMA in Robotic Communications/Autonomous Networks}
Autonomous robotics, which have brought tremendous changes in various socio-economic aspects in our society, have attracted significant attention in the wireless communications field. Autonomous robots are classified into three categories: aerospace robots (e.g., airship/airplane, HAP, and UAV); ground robots (e.g., autonomous vehicle, smart home robots, and mobile robots); marine robots (e.g., unmanned ship, unmanned submarine, underwater robots). Before fully reaping the benefits of autonomous robotics, challenges such as delay sensitivity and collision tolerance have to be tackled. With the aid of NOMA, the communication signal is designed to be ultra-reliable and low-latency, which fits the requirement of autonomous robotic systems. Additionally, since robots often serve in dynamic scenarios, the channel conditions of robots are varying at each timeslot, which also improves the design flexibility of NOMA-based networks. By dynamically assigning the robots and other mobile users with heterogeneous mobility states into the same NOMA cluster to activate hybrid-SIC-based OTSF-NOMA schemes, the received SINR of both robots and mobile users can be enhanced. Thus, NOMA can be helpful for the efficiency and safety of autonomous robotics.
\subsection{NOMA in VR/AR Multi-Layer Video Transmission}

VR and AR, which provide a better experience for users because of their immersive scenes, have been recognized as one of the key scenarios in 6G networks. In the conventional OMA schemes, resources are allocated by blocks of a fixed size as it is non-trivial to divide them into small sizes, which means that OMA users have to wait for access to resources. However, in VR and AR multi-layer video transmission networks, services to users are often delay-sensitive while high transmit rate and reliable communications are highly demanded, which highlights the importance of high spectral efficiency in wireless communication. Since hybrid-SIC enabled NOMA is capable of cancelling the outage probability error floor of primary users for providing a reliable transmission, applying hybrid-SIC enabled NOMA in VR and AR multi-layer video transmission networks is necessary.

\begin{table*}[t!]
  \footnotesize
  \begin{center}
\begin{threeparttable}
\caption{Pros and Cons of Machine Learning Algorithms in NOMA-based Networks}
\label{pros}
\begin{tabular}{|c||c||c|}
\hline
\centering
  & Conventional solutions &  ML-empowered solutions  \\
\hline
Representative algorithms& \tabincell{c}{Iterative algorithms, matching theory, \\
game theory, alternating algorithms }  &   \tabincell{c}{DL, QL algorithm, DQN algorithm, \\DDPG algorithm }\\
\hline
\tabincell{c}{Constant vs \\ Time-variant QoS requirement}& Ignore user behavior/peculiarity &  User behavior/peculiarity is considered \\
\hline
\tabincell{c}{Static user vs\\ Heterogeneous user mobility} & Cannot adapt to the dynamic environment &  \tabincell{c}{Rapidly adapting to the dynamic environment\\ (Dynamic NOMA user re-clustering and resource allocation)}    \\
\hline
\tabincell{c}{Short-term vs\\ Long-term resource allocation} & Focus on benefits of the current timeslot &   Can incorporate farsighted system evolution   \\
\hline
\tabincell{c}{Non-interactive vs \\Interactive with environment}& \tabincell{c}{No learning capability \\ or interactive capability} &   \tabincell{c}{Learn from the environment \\and real-time feedback of users }  \\
\hline
Optimal vs Sub-optimal& Optimality can be proved &   \tabincell{c}{Optimality cannot be theoretically proved \\or strictly guaranteed}   \\
\hline
\end{tabular}
\begin{tablenotes}
\item[1] QL represents Q-learning; DQN is short for Deep Q-Network; DDPG is the acronym for deep deterministic policy gradient base algorithms.
\end{tablenotes}
\end{threeparttable}
 \end{center}
\end{table*}
\subsection{NOMA in E-Health Systems}
Due to its benefits of supporting enhanced spectrum efficiency and massive connectivity, NOMA has attracted increased attention for emerging remote e-Health services in recent years. Besides the power-domain NOMA, the QoS-based NOMA scheme can also be applied in communication-aware E-health systems. The QoS-based SIC scheme can be applied to first detect the inquired user's (delay-sensitive) signal and then decode the monitored user's (delay-tolerant) information. Thus, the precious communication resources can be efficiently utilized. Since the number of smart devices (e.g., smart watches) in communication-aware E-health systems have been growing exponentially, hybrid-SIC enabled NOMA is helpful for providing massive connectivity with strict QoS requirements in E-health systems and providing additional access channels for offloading heavy computation tasks to edge devices or edge servers.
\section{New Tools: Machine Learning Empowered NOMA-based Networks}
Since scenarios towards 6G networks are heterogeneous, dynamic, and more complex, new approaches are needed to tackle more challenging problems in these scenarios~\cite{Saad20206G}. Most conventional approaches adopted in the current research contributions have a limited number of controllable variables and fail to optimise Markovian problems efficiently. Most conventional approaches adopted in the current research contributions have a limited number of controllable variables and fail to optimise Markovian problems efficiently. Moreover, comparing with conventional NOMA scheme with single-objective (CSI or QoS) user clustering and power allocation, the hybrid SIC as discussed in Section II introduces more controllable variables as it has multi-objective (CSI and QoS) user clustering and power allocation. Therefore, machine learning becomes a competitive candidate for designing NGMA. In contrast to conventional algorithms, existing AI-based algorithms are capable of addressing problems in a more general framework with a large number of controllable variables and temporal correlated events. The general advantages/disadvantages of applying machine learning approaches in NOMA-based networks are listed in Table~\ref{pros}. Additionally, machine learning (ML) empowers NOMA-based networks to realize the interaction between BSs and the dynamic environment. More importantly, in the conventional solutions, the control policy aims for instantaneously achieving the current benefits for networks without considering the long-term network evolution, which is the objective of ML algorithms. This section introduces the role of ML in NOMA-based networks.
\vspace{-0.3 cm}
\subsection{Reinforcement Learning for NOMA-based Networks}
Reinforcement learning (RL) models enable BSs/access points (APs) to learn from the real-time feedback of dynamic/uncertain environment and mobile users, as well as from their historical experiences~\cite{Letaief20196G}. Thus, in RL-empowered NOMA-based networks, BSs/APs are capable of rapidly adapting their control policy to the dynamic environment and keep on improving their decision-making ability. Application examples of RL in NOMA-based networks include:
\begin{itemize}
\item \textbf{Dynamic User Clustering/Pairing:} The optimal policies for both CSI-based and QoS-based user clustering in NOMA-based networks are challenging, especially when user mobility is considered. By adopting RL-based algorithms for user re-clustering, the performance of such networks is improved in dynamic scenarios.
\item \textbf{Long-term Resource Allocation:} RL aims at achieving long-term benefits by deciding the resource allocation (e.g., power resource, caching resource, computing resource, spectrum resource) policy in NOMA-based networks. RL-based algorithms are capable of outperforming the conventional algorithms in dynamic scenarios by interacting with the environment. However, since the optimality of RL-based algorithms cannot be strictly guaranteed, their superiority is reduced in some simple static scenarios.
\end{itemize}

RL models are suitable for Markovian problems, which means that the formulated problem in NOMA-based networks has to be reformulated as a Markovian one.

\subsection{Deep Learning for NOMA-based Networks}
Deep learning (DL), which uses cascaded neural network layers to extract features from the input data automatically and make a decision, has shown great potentials to revolutionize NOMA-based networks~\cite{Mao2018DL}. Application examples of DL in such networks include:
\begin{itemize}
\item \textbf{CSI Acquisition/Channel Estimation:} As introduced above, CSI-based SIC is a fundamental principle of power-domain NOMA. However, acquiring CSI is challenging, especially when integrating NOMA with other techniques such as MIMO and RIS. With the aid of DL, the CSI can be acquired automatically by extensive training of the input data using existing channel models. Additionally, DL can also be applied for detecting/estimating the dynamically fluctuating channel.
\item \textbf{Resource Allocation:} Since the performance of NOMA-based networks is significantly affected by the resource allocation policy, resource management is a principal problem in such networks. However, resource allocation under dynamic channel conditions is challenging. Thus, DL is attractive for dynamic resource allocation in NOMA-based networks, including power allocation, subchannel assignment, and subcarrier assignment.
\end{itemize}

DL-based algorithms are also attractive for solving other problems in such networks, such as signal constellation design, signal detection, and signal decoding. Although DL has been proved to be efficient and to provide effective performance improvement, the efficient design of DL architecture (e.g., number of layers, the connections between layers) to reduce the computational complexity is still an open problem.

\subsection{Other Machine Learning for NOMA-based Networks}
In addition to DL-based and RL-based based algorithms, unsupervised learning algorithms can be also adopted for user clustering. By exploiting the correlation feature of users' channel responses, the K-means algorithm can be adopted for NOMA user clustering. By learning the inherent structures and correlations of users, the expectation maximization algorithm can be used for user clustering in both fixed and dynamic user scenarios. Federated learning (FL), which relies on directly training statistical models on remote devices at the edge in distributed networks, is very attractive to preserve users data privacy in wireless networks~\cite{Niknam2020FL}. NOMA is able to reduce the aggregation latency in FL model updating by providing multiple access in the same channel. In FL-enabled NOMA-based networks, BSs act as distributed learners, which train their generated data and transfer their local model parameters instead of the raw training dataset to an aggregating unit. Moreover, based on well-trained DL and DRL, transfer learning can be used to update the full/partial neural network parameters for fast convergence at a new environment with similar tasks but limited data or training time.

\section{Conclusion and Outlook}
In this article, we described our view on the advances of NOMA for 6G based on the ``one principle plus four new'' concept. As a core principle, the hybrid SIC generalized CSI-based decoding order to a joint CSI- and QoS-based one, which is able to remove the conventional outage probability error floor for joining users. Building on this principle, four new directions, including massive connectivity, the integration with other physical layer techniques, the application to heterogeneous scenarios, and the AI enhancement of NOMA, were presented in detail. These discussions demonstrated that NOMA has tremendous benefits for the NGMA family, especially for networks with distinct QoS requirements.

\bibliographystyle{IEEEtran}

\begin{IEEEbiographynophoto} {Yuanwei Liu}
[SM'19] (yuanwei.liu@qmul.ac.uk) is a senior lecturer (associate professor) at Queen Mary University of London. His research interests include 5G/6G networks, NOMA, RIS, UAV communications, and machine learning. He is a Web of Science Highly Cited Researcher 2021. He is currently a Senior Editor of IEEE Communications Letters, and an Editor of IEEE Transactions on Wireless Communications and IEEE Transactions on Communications. He received the IEEE ComSoc Outstanding Young Researcher Award for EMEA in 2020. He received the 2020 IEEE Signal Processing and Computing for Communications Technical Early Achievement Award and the IEEE Communication Theory Technical Committee 2021 Early Achievement Award. He received the IEEE ComSoc Young Professional Outstanding Nominee Award in 2021.
\end{IEEEbiographynophoto}
\vspace{-10mm}
\begin{IEEEbiographynophoto} {Wenqiang Yi}
[M'20] (w.yi@qmul.ac.uk) has been a postdoctoral researcher at Queen Mary University of London, since 2020. His research interests include NGMA, mmWave communications, RIS, stochastic geometry, and reinforcement learning. He is the exemplary reviewer of IEEE Transactions on Communication and IEEE Communications Letter in 2019 and 2020. He serves as the Secretary of the SIG on Signal Processing Techniques for Next Generation Multiple Access (NGMA) by the SPCC Technical Committee.
\end{IEEEbiographynophoto}
\vspace{-10mm}
\begin{IEEEbiographynophoto}{Zhiguo Ding}
[F'20] (zhiguo.ding@manchester.ac.uk) is currently a Professor at the University of Manchester. From Sept. 2012 to Sept. 2022, he has also been an academic visitor in Princeton University. Dr Ding' research interests are B5G networks and statistical signal processing. He is an area editor for IEEE OJ-COMS and an Editor for IEEE TVT, and served as an editor for IEEE TCOM, WCL and IEEE CL. He received 2018 IEEE COMSOC Heinrich Hertz Award, IEEE VTS Jack Neubauer Memorial Award, Friedrich Wilhelm Bessel Research Award 2020, and IEEE SPCC Technical Recognition Award 2021. He is a Distinguished Lecturer of IEEE ComSoc, and a Web of Science Highly Cited Researcher in two categories 2021.
\end{IEEEbiographynophoto}
\vspace{-10mm}
\begin{IEEEbiographynophoto} {Xiao Liu}
(x.liu@qmul.ac.uk) received the Ph.D. degree from Queen Mary University of London in 2021. His research interests include UAV aided networks, machine learning, NOMA techniques, RIS. He is the exemplary reviewer of IEEE Transactions on Communication and IEEE Communications Letter, 2019.
\end{IEEEbiographynophoto}
\vspace{-10mm}
\begin{IEEEbiographynophoto} {Octavia A. Dobre}
[Fellow, IEEE] (odobre@mun.ca) is a Professor and Research Chair at Memorial University, Canada. Her research interests include technologies for beyond 5G, as well as optical and underwater communications. She published over 400 referred papers in these areas. Dr. Dobre serves as the Editor-in-Chief (EiC) of the IEEE Open Journal of the Communications Society. She was the EiC of the IEEE Communications Letters, a senior editor and an editor with prestigious journals, as well as General Chair and Technical Co-Chair of flagship conferences in her area of expertise. Dr. Dobre is the recipient of diverse awards, such as Best Paper Awards at IEEE ICC, IEEE Globecom, IEEE WCNC, and IEEE PIMRC conferences. She is an elected member of the European Academy of Sciences and Arts, a fellow of the Engineering Institute of Canada, and a fellow of the Canadian Academy of Engineering.
\end{IEEEbiographynophoto}
\vspace{-10mm}
\begin{IEEEbiographynophoto} {Naofal Al-Dhahir}
(aldhahir@utdallas.edu) is Erik Jonsson Distinguished Professor \& ECE Dept. Associate Head at UT-Dallas. He earned his PhD degree from Stanford University and was a principal member of technical staff at GE Research Center and AT\&T Shannon Laboratory from 1994 to 2003.  He is co-inventor of 43 issued patents, co-author of about 500 papers and co-recipient of 5 IEEE best paper awards. He is an IEEE Fellow, received 2019 IEEE SPCC technical recognition award and 2021 Qualcomm faculty award. He served as Editor-in-Chief of IEEE Transactions on Communications from Jan. 2016 to Dec. 2019.  He is a Fellow of the National Academy of Inventors.
\end{IEEEbiographynophoto}
 \end{document}